\documentstyle[preprint,aps,eqsecnum,epsfig]{revtex}

\newcommand{\beq}{\begin{equation}}
\newcommand{\eeq}{\end{equation}}
\newcommand{\beqs}{\begin{eqnarray}}
\newcommand{\eeqs}{\end{eqnarray}}

\def\mathcal{\cal}
\def\mathrm{\rm}
\def\textbf{\bf}
\def\textit{\it}
\def\lesssim{\ \rlap{\raise 3pt \hbox{$<$}}{\lower 3pt \hbox{$\sim$}}\ }
\def\gtrsim{\ \rlap{\raise 3pt \hbox{$>$}}{\lower 3pt \hbox{$\sim$}}\ }

\def\ifmath#1{\relax\ifmmode #1\else $#1$\fi}
\def\ifmath{}

\begin{document}
\draft

{\tighten
\preprint{\vbox{\hbox{WIS-99/25/JUL-DPP}
                \hbox{hep-ph/9907205}}}
\title{~\\~\\Implications of Neutrino Masses on 
the $K_{L}\rightarrow\pi^{0}\nu\bar{\nu}$ Decay}

\author{Gilad Perez}
\address{ \vbox{\vskip 0.truecm}
  Department of Particle Physics,
  Weizmann Institute of Science, Rehovot 76100, Israel }
\maketitle

\begin{abstract}
We calculate the different contributions to the decay 
$K_{L}\rightarrow \pi ^{0}\nu \bar{\nu }$ that arise if the 
neutrinos are massive. In spite of a chiral enhancement 
factor, we find that these contributions are negligibly 
small. Compared to the CP violating leading contributions, 
the CP conserving contributions related to Dirac masses 
are suppressed by a factor of order 
$(m_K m_\nu/m_W^2)^2\lesssim10^{-12}$, and those related
to Majorana masses are suppressed by a factor of order
$(\alpha_W m_K m_s^2 m_\nu/m_W^4)^2\lesssim10^{-29}$.
With lepton flavor mixing we find new contributions with
a single CP violating coupling leading to a final CP even 
state or with two CP violating couplings leading to a 
final CP odd state. These contributions can be of the 
order of the Standard Model CP conserving contributions to 
the flavor diagonal modes.
\end{abstract} 
} 

\newpage

\section{Introduction}
The decay $K_{L}\rightarrow \pi ^{0}\nu \bar{\nu}$ is known to be 
a purely CP violating (CPV) process to a very good approximation 
\cite{LIT} and subject to a clean theoretical interpretation. 
Within the standard model (SM), it will determine the physical CPV 
phase ($\eta$) \cite{Bur}. Beyond the SM, it will probe new sources 
of CPV \cite{GN}. 

The general argument that the CP conserving (CPC) contributions are small
is based on the chiral expansion \cite{BaGl}. Within the SM, these
contributions have been calculated and found to be indeed negligible 
\cite{BI}. 

If, however, neutrinos were massive then a scalar four fermion operator 
would be allowed. Such an operator is CPC and would be the lowest order 
in the chiral expansion. On the other hand, it is suppressed by the 
small neutrino mass. In addition, if neutrinos are massive then flavor 
mixing is possible. A $\pi^0\nu_i\bar\nu_j$ final state with $i\neq j$,
which is not a CP eigenstate, is allowed \cite{GN}. 
This would give rise to additional contributions with a single or 
double CPV coupling leading to a CP even or CP odd final states 
respectively. 
In this work we calculate the different contributions 
related to neutrino masses and mixing and find 
whether they can be significant.

The models that we consider here are minimal extensions of the SM.
We take the SM Lagrangian and add to it the Yukawa interactions that 
are necessary to induce neutrino masses. We assume that all other
effects of new physics related to neutrinos are negligibly small and do
not affect the $K_{L}\rightarrow \pi ^{0}\nu \bar{\nu }$ decay.
We consider two types of neutrino masses: 
\begin{itemize}
\item[1.] Dirac masses, which require the existence of right handed 
neutrinos. Lepton number is conserved but there is
no understanding of why neutrinos are lighter than charged fermions.
We assume that the right-handed neutrinos are only involved in
Yukawa interactions. In particular, if they have gauge interactions
beyond the SM, we assume that these interactions take place
at a high enough energy scale that they effectively decouple.
\item[2.] Majorana masses, which require that the SM is taken to
be a low energy effective theory only. Lepton number is violated and
there is a natural explanation of the lightness of neutrinos \cite{See}.
We assume that the scale of lepton number violation is high enough
that any other effects of physics at that scale effectively decouple.
\end{itemize}
We study Dirac neutrinos in section II and Majorana neutrinos in section
III. The effects of lepton flavor mixing
are analyzed in section IV. A summary of our conclusions is given
in section V.

\section{Massive Dirac Neutrinos}
In the presence of neutrino Dirac masses, new CPC operators could 
appear in the effective Hamiltonian: A scalar operator,
\begin{equation}
{\mathcal{H}}_{eff}^{scal}=(\bar{s}d)(\bar{\nu}\nu ),  \label{hmdirac}
\end{equation}
and a tensor operator,
${\mathcal{H}}_{eff}^{ten}=(\bar{s}\sigma ^{\mu \nu }\gamma ^{5}d)(\bar{\nu }
\sigma _{\mu \nu }\gamma _{5}\nu ).$
The tensor operator contribution is expected to be smaller than the scalar 
one \cite{KTPNT}, and we do not consider it any further here.

\subsection{The model}
We consider an extension of the SM where we add three right-handed (singlet)
neutrinos and impose lepton number conservation. The right-handed neutrinos
have no gauge interactions, but they have Yukawa couplings:
\begin{equation}
{\mathcal{L}}_{Y}=f_{mn}\overline{L_{Li}^{m}}\nu _{R}^{n}\tau
_{2}^{ij}\Phi_{j}^{\ast }+{\mathrm{h.c.}},  \label{Lnifi}
\end{equation}
where $L_{L}$ is a left-handed lepton doublet, $\Phi$ is the Higgs
doublet and $\nu _{R}$ is a right-handed neutrino singlet; $m,n$ are 
generation indices and $i,j$ are SU(2) ones.
Similarly to the SM, the only source of flavor changing couplings (in the mass
basis) are the charged current interactions, 
while the Higgs couplings are diagonal. 
In this section we consider
only outgoing neutrinos which carry the same flavor. 
CPV and mixing effects in the lepton sector will 
be treated separately in section IV. 

\subsection{The Decay Rate Calculation}
We calculate the amplitude in the full theory and without QCD corrections.
Neglecting the long distance contributions is justified since we find
that the dominant contribution arises from loop momenta 
$k\sim M_{W}$. The omission of QCD corrections does not affect our 
conclusions.

The diagrams that generate the scalar operator (\ref{hmdirac})
are of two types. First, there are the SM box diagrams 
drawn in fig. \ref{fig4box}. Second, there is a 
neutral Higgs mediated penguin diagram, related to the Yukawa 
interaction of eq. (\ref{Lnifi}). Replacing the SM loop contribution 
with an effective $sdH$ coupling \cite{WY}, we can consider it to be an 
``effective tree'' diagram drawn in fig. \ref{figsdHnini}.

\subsubsection{The box diagram contributions}
In contrast to the calculation of the CPV contributions from
box diagrams, external momenta should not be neglected in our calculation. 
The external momentum expansion will produce a CPC operator of the form 
$\overline{d}p_{s}^{\mu }\gamma _{\mu }s$ $\sim \overline{d}m_{s}s$,
which is comparable with the other contributions. The calculation is, 
however, simplified by the fact that the dominant terms have at most 
a linear dependence on the external momentum.
Contributions suppressed by a factor of order $(m_{i}m_{j})/M_{W}^{2}$
(where $i,j=u,d,\nu $) are subdominant and we neglect them.

\begin{figure}[hbct]
\begin{center}
\mbox{\epsfig{figure=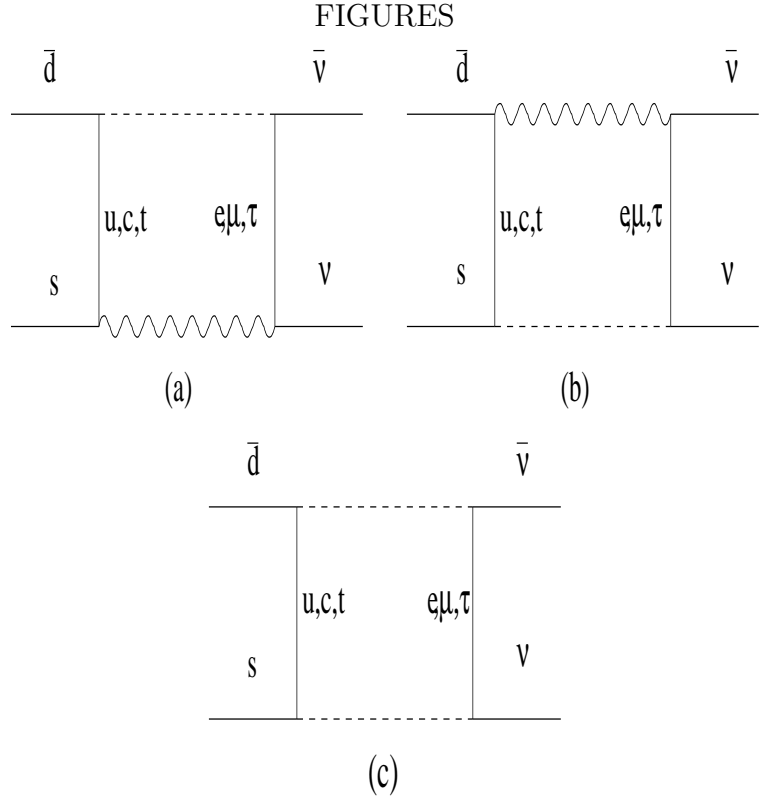,angle=0,width=10cm,height=10cm}}
\end{center}
\caption{The Standard Model box diagrams.}
\label{fig4box}
\end{figure}

Summing the dominant contributions from the box diagrams
(fig. \ref{fig4box}), we get: 
\begin{equation}
{\mathcal{H}}_{CPC}^{^{\ref{fig4box}}}=\sum_{i,k,\ell}\frac{G_{F}}{\sqrt{2}}
\frac{\alpha }{2\pi \sin ^{2}\Theta _{W}}\lambda _{i}
 \widetilde{V}_{\ell k}^{\ast }\widetilde{V}_{\ell k}
(\bar{d}s)(\overline{\nu }_{\ell }\nu _{\ell })
X_{B}(x_{i})+\mathrm{h.c.}~,  \label{Hboxtot}
\end{equation}
with $i,k,\ell$ the flavor indices of the internal quark, internal lepton
and external neutrinos respectively;  
$\widetilde{V} $ is the MNS lepton mixing matrix \cite{MNS};
\begin{eqnarray}
X_{B}(x_{i})=\frac{m_{\nu }m_{s}}{2M_{W}^{2}}\left\{ \frac{(-x_{i}+1
+x_{i}\ln (x_{i}))}{(x_{i}-1)^{2}}-\frac{1}{2}\frac{
x_{i}(-2x_{i}+\ln (x_{i})x_{i}+\ln (x_{i})+2)}{(-1+x_{i})^{3}}\right\}  
\label{Xtot} 
\end{eqnarray} 
and $x_{i}=(m_{i}/M_{W})^{2}$. Since the top quark contribution 
is dominant, (\ref{Xtot}) can be simplified: 
\begin{equation}
\sum_{i}\lambda_{i}X_{B}(x_{i})\approx \lambda_t
\frac{m_{\nu }m_{s}}{M_{W}^{2}}\frac{x_{t}}{2(x_{t}-1)^{2}}
\left( 2-x_{t}+\ln (x_{t})\frac{x_{t}-3}{2(x_{t}-1)}\right) ~
.  \label{XtotGIM}
\end{equation} 
Moreover since $X_{B}$ (\ref{Xtot}) is to leading order $m_\ell$ 
independent and by the virtue of $\widetilde{V}$ unitarity, the 
sum over $k$ is trivial: 
\begin{equation}
\sum_{k}\widetilde{V}_{\ell k}^{\ast }\widetilde{V}_{\ell k}
X_{B}(x_{i})=X_{B}(x_{i}) .
\end{equation}
\label{MNSU}
We learn that CPC contributions from box diagrams 
are suppressed by a factor of ${\cal O}(m_{\nu}m_{s}/M_{W}^{2})$.

\subsubsection{The Higgs-mediated contribution}
\begin{figure}[hbct]
\begin{center}
\mbox{\epsfig{figure=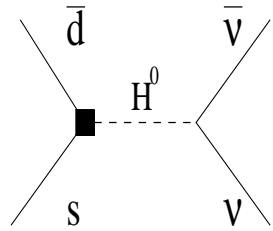,angle=0,width=3.5cm,height=3cm}}
\end{center}
\caption{The Higgs-mediated diagram for the case of Dirac masses.}
\label{figsdHnini}
\end{figure}

We calculate the Higgs-mediated diagram of fig. \ref{figsdHnini}. 
The effective $s\overline{d}H$ coupling $\Gamma$ arising from the
SM loop (represented as a square in fig. \ref{figsdHnini}) 
is given by \cite{WY}: 
\begin{equation}
\Gamma\equiv-\lambda _{t}\frac{g^{3}}{128\pi^{2}}
\frac{m_{t}^{2}m_{s}}{M_{W}^{3}} \left(\frac{3}{2}
+\frac{m_{H}^2}{M_{W}^{2}}f_2(x_t)
\right)(1+\gamma^5),\label{effSDH}
\end{equation}
where 
\begin{equation}
f_2(x)\equiv \frac{x}{2(1-x)^2}\left(-\frac{x}{1-x}\ln x+\frac{2}{1-x}\ln x
-\frac{1}{2}-\frac{3}{2x}\right) \label{f_x}.
\end{equation}
Since the coupling in eq. (\ref{effSDH}) is by itself proportional to 
$\frac{m_{s}}{M_{W}}$, we can safely neglect the external momentum. 
Thus the amplitude of fig. \ref{figsdHnini} is given by: 
\begin{equation}
i{\mathcal{T}}^{\ref{figsdHnini}}=-\sum_{\ell}\frac{G_{F}}{\sqrt{2}}
\frac{\alpha }{(2\pi )\sin ^{2}\Theta _{W}}\frac{i}{2}
\frac{m_{t}^{2}m_{s}m_{\nu}}{M_{H}^{2}M_{W}^{2}}
\lambda _{t}\left(\frac{3}{2}+\frac{m_{H}^2}{M_{W}^{2}}f_2(x_t) \right)
\,(\bar{d}s_{R})(\overline{\nu }_{\ell}\nu _{\ell })~,
\label{Teffpen}
\end{equation}

\subsubsection{The rate}

Adding the contributions of the Higgs-penguin and the box diagrams 
(eqs. (\ref{Hboxtot}) and (\ref{Teffpen})), we get: 
\begin{equation}
{\mathcal{H}}_{CPC}^{Dir}=\sum_{\ell}\frac{G_{F}}{\sqrt{2}}\frac{\alpha}
{2\pi \sin ^{2}\Theta _{W}}\lambda _{t}  
\ (\bar{d}s)(\overline{\nu }_{\ell}\nu _{\ell })
X^{Dir}+\mathrm{h.c.}~,  \label{HDCPC}
\end{equation}
where 
\begin{equation}
X^{Dir}\approx \frac{m_{\nu }m_{s}}{M_{W}^{2}}\frac{x_{t}}{2}
\left[ \frac{1}{(x_{t}-1)^{2}}
\left( 2-x_{t}+\ln (x_{t})\frac{x_{t}-3}{2(x_{t}-1)}\right) 
-\frac{1}{2}\frac{M_{W}^{2}}{M_{H}^{2}}
\left(\frac{3}{2}+ \frac{m_{H}^2}{M_{W}^{2}}f_2(x_t) \right)\right] ~ .
\label{Xdirac}
\end{equation}
The ratio between the scalar hadronic matrix element and the 
$V-A$ hadronic matrix element (related to the leading CPV contribution) 
is found via the use of the equation of motion 
\cite{KM}:
\begin{equation}
\frac{\mid \langle \pi ^{0}|\bar{s}d|K^{0}\rangle \mid }{\mid 
\langle \pi ^{0}|(\bar{s}d) _{V-A}|K^{0}\rangle \mid }\sim \frac{M_{K}}
{m_{s}}\approx 3~\, .
\label{RdCPCCPV}
\end{equation}
Using eqs. (\ref{HDCPC}) and (\ref{RdCPCCPV}), we can now compare the 
leading CPV rate \cite{IL} to the CPC one with massive Dirac neutrinos: 
\begin{equation}
R_{CPV}^{Dir}=\frac{\Gamma _{CPC}^{Dir}(K_{L} 
\rightarrow \pi ^{0} \nu \bar{\nu})}{\Gamma _{CPV}
(K_{L}\rightarrow \pi ^{0}\nu 
\bar{\nu})}\sim \left( \frac{M_{K}}{M_{W}}\frac{m_{\nu }}{M_{W}}\right)
^{2}\approx 10^{-12}\left[ \frac{m_{\nu }}{10~ MeV}
\right] ^{2}~ .  \label{RGdCPCCPV}
\end{equation}
The direct experimental upper bound \cite{EPJ},
$m_{\nu _{\tau }}$ $\leq 18.2~MeV$, implies that indeed the CPC 
contribution from neutrino Dirac masses is suppressed by at least 
twelve orders of magnitude compared to the CPV contribution of the SM. 
However, a significantly lower upper bound ($m_{\nu }$ $\lesssim
O(10~eV)$) is found from astrophysical and cosmological reasoning \cite{Neu}, 
so that very likely $R _{CPV}^{Dir}\lesssim 10^{-24}$.
We learn that the addition of a small Dirac mass to the SM neutrinos does not
change the statement that the $(K_{L}\rightarrow \pi ^{0}\nu \bar{\nu})$ 
decay is purely CPV to an excellent approximation. 

\section{Massive Majorana Neutrinos}
In the presence of neutrino Majorana masses, a new scalar, 
CPC operator appears in the effective Hamiltonian: 
\begin{equation}
{\mathcal{H}}_{eff}^{Maj}=(\bar{s}d)(\nu\nu ).  \label{hmmajor}
\end{equation}

\subsection{The Model}
We consider the SM as a low energy effective theory, that
is we allow for non-renormalizable terms. Neutrino Majorana masses 
are induced via the following dimension-five terms:
\begin{equation}
{\mathcal{L}}_{\nu \nu }={\frac{1}{2}}\left( \frac{f_{mn}}{M}\right)
(L_{Li}^{m})_{\alpha }\sigma _{\alpha \beta }^{2}\tau _{ir}^{2}\tau
_{rj}^{t}(L_{Lj}^{n})_{\beta }\Phi _{k}\tau _{ks}^{2}\tau _{sl}^{t}\Phi _{l}
+\mathrm{h.c.}~ ,  \label{majmassterm}
\end{equation}
where  $m,n$ the generation indices, $i,j,k,l,r,s,t$ are 
SU(2) indices and $\alpha ,\beta$ are spinor indices.
In this section we assume again flavor diagonality.  Flavor mixing 
is discussed in the next section.  

The interaction term (\ref{majmassterm}) induces neutrino masses,
$(M^{Maj}_\nu)_{mn}\sim f_{mn}{\langle\Phi^0\rangle^2\over M}$.
In addition, it generates a new CPC contribution to the 
$K_{L}\rightarrow \pi ^{0}\nu \bar{\nu}$ decay. 
The diagram which leads to the CPC contribution is shown in 
fig. \ref{figmaj}.
The effective ${\nu }_{{L}}^{T}\sigma ^{2}\nu _{L}HH$ coupling of the 
diagram in fig. \ref{figmaj} can be written in
terms of the neutrino mass: 
\begin{equation}
-\frac{f_{ii}}{M}=-g^{2}\frac{m_{\nu _{i}}^{Maj}}{2M_{W}^{2}}~ ,
\label{effcoup}
\end{equation}
with $i=1,2,3$ \,.
Note that the diagram includes the effective $s\overline{d}H$ coupling
given in eq. (\ref{effSDH}).

\subsection{The Decay Rate Calculation}
\subsubsection{The amplitude}
\begin{figure}[hbct]
\begin{center}
\mbox{\epsfig{figure=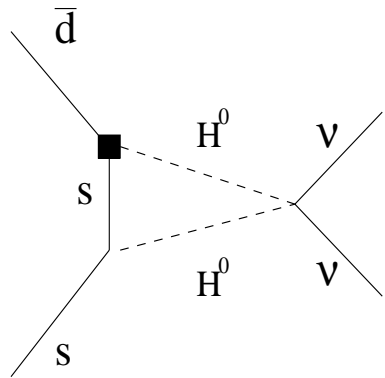,angle=0,width=5cm,height=5cm}}
\end{center}
\caption{The Majorana case CPC diagrams.}
\label{figmaj}
\end{figure}

The diagram in fig. \ref{figmaj} is calculated using standard
techniques. The fact that we have two suppressed vertices
allows us to neglect the external momenta. We again neglect QCD corrections.
The short distance contribution is: 
\begin{eqnarray}
i{\mathcal{T}}^{\ref{figmaj}} &=&-\sum_{i,k,\ell}
\frac{G_{F}}{\sqrt{2}}
\frac{\alpha }{(2\pi )\sin ^{2}\Theta _{W}}(M_{W}^{2}16\pi ^{2})\lambda _{t}
\frac{m_{t}^{2}m_{s}^{2}m_{\nu }^{Maj}}{M_{W}^{6}} 
\frac{g^{2}}{128\pi^{2}} \left(\frac{3}{2}
+\frac{m_{H}^2}{M_{W}^{2}}f_2(x_t)\right)
\label{Tshmaj} \\
&&\times \int_{\mu }^{\infty }\frac{d^{4}k}{(2\pi )^{4}}\frac{i}{
(k^{2}-M_{H}^{2})^{2}}\left[ \bar{d}_{R}i\frac{\gamma \cdot k+m_{s}}{
k^{2}-m_{s}^{2}}s\right] \left( ({\nu }_{\ell L})^{T}\sigma ^{2}\nu
_{\ell L}\right) ~ .  \nonumber
\end{eqnarray}
Thus we get the following CPC operator: 
\begin{equation}
{\mathcal{H}}_{CPC1}^{\ref{figmaj}}\approx \sum_{i,\ell}
\frac{G_{F}}{\sqrt{2}}
\frac{\alpha }{2\pi \sin ^{2}\Theta _{W}}\lambda _{i}X_{Maj1}\ (\bar{d}
s_{R})\left( ({\nu }_{\ell L})^{T}\sigma ^{2}\nu _{\ell L}\right) +
\mathrm{h.c.}~,  \label{hshmaj}
\end{equation}
where 
\begin{equation}
X_{Maj1}=\frac{1}{4}I_{Maj1}g^{2}\frac{ m_{t}^{2}m_{s}^{3}m_{\nu }^{Maj}} 
{M_{H}^{2}M_{W}^{4}} 
\left(\frac{3}{2}+\frac{m_{H}^2}{M_{W}^{2}}f_2(x_t)\right)
\label{Xshmaj}
\end{equation}
and, defining $y_{s}\equiv\frac{m_{s}^{2}}{M_{H}^{2}}$,
\begin{equation}
I_{Maj1}=\int_{\mu /M_{H}}^{\infty }\frac{-id^{4}k}{(2\pi )^{4}}
\frac{1}{(k^{2}-1)^{2}}\frac{1}{k^{2}-y_{s}}\approx \frac{1+2y_{s}
\ln(\mu/M_{H})}{32\pi^{2}}~ .  
\label{Imaj1}
\end{equation}
Note that the long distance contributions (namely 
the contributions from $k\lesssim \Lambda _{QCD}$ in \ref{Imaj1}), 
are logarithmic and suppressed by $y_s \sim 10^{-6}$. 
Therefore they are subdominant and neglected.

The hadronic matrix element is the same as in the case of the
Dirac neutrinos. Its ratio with the leading CPV hadronic operator 
was presented above (\ref{RdCPCCPV}).

\subsubsection{The rate}
We can now compare the Majorana mass contribution to the CPC rate 
with the leading CPV \cite{IL} one. Using (\ref{RdCPCCPV})
and (\ref{hshmaj}), we find: 
\begin{eqnarray}
R_{CPV}^{Maj} &\equiv& \frac{\Gamma _{CPC}^{Maj\ }(K_{L}
\rightarrow \pi ^{0} \nu \bar{\nu})}{\Gamma_{CPV}^{SM}
(K_{L}\rightarrow \pi ^{0}
\nu \bar{\nu})}\sim \left( \frac{3\alpha }{
64\pi \sin ^{2}\Theta _{W}}\frac{m_{t}^{2}}{M_{H}^{2}}
\frac{m_{s}^{3}}{M_{W}^{3}}{\frac{m_{\nu }^{Maj}}{M_{W}}} 
\frac{M_{K}}{m_{s}}\right) ^{2} \label{RGmCPCCPV} \\ 
&\approx& 10^{-29}\left[ \frac{m_{\nu }}{10~
{MeV}}\right] ^{2}\left[ \frac{100~{GeV}} {M_{H}}
\right] ^{4}~ \nonumber . 
\end{eqnarray}
We learn that the CPC contribution
from neutrino Majorana masses is suppressed by at least twenty nine 
orders of magnitude compared to the CPV contribution of the SM.
Since, very likely, $m_{\nu _{\tau }}\lesssim 10~eV$ \cite{Neu}, 
we expect $R_{CPV}^{Maj}\lesssim 10^{-41}.$

The strong suppression of the massive Majorana CPC amplitude is a result of
four factors: First, the CPC operator ${\nu }_{{L}}^{T}\sigma ^{2}\nu _{L}$
is created via the ${\nu }_{L}^{T}\sigma ^{2}\nu _{L}HH$ vertex. The
vertex carries suppression factor $\frac{m_{\nu }^{Maj}}{M_{W}}$. A second
factor, $\frac{3g^{2}}{128\pi ^{2}}\frac{m_{s}}{M_{W}}$, arises from the
flavor diagonality of the tree level Yukawa coupling: it is the loop
suppression of the $s\overline{d}H$ coupling.
The third factor, $\frac{m_{s}}{M_{W}}$, arises from the
$s\overline{s}H$ Yukawa coupling. The fourth one, $\frac{m_{s}}{
M_{W}}$, arises due to the required helicity flip along the internal $s$
quark propagator.

\section{Lepton Flavor Mixing}

In the presence of masses, the neutrino mass basis is in general not 
equal to the interaction basis, leading to lepton flavor mixing.
The effects of mixing on the different contribution to the 
$K_{L}\rightarrow \pi ^{0}\nu \bar{\nu}$  decay are analyzed below.
We first investigate the effects of mixing on the contributions related
to the scalar operator. 
Then we point out that the contributions related to $V-A$ operators can 
lead, through two insertions of CP violating vertices, to CP odd 
final states.

\subsection{Scalar Operators}
\subsubsection{Dirac neutrinos}
With flavor mixing, the effective Hamiltonian which governs the decay 
acquires the following form: 
\begin{equation}
\label{Hdirmix}
{\mathcal{H}}_{mix}^{Dir}=\sum _{j,k,\ell }\frac{G_{F}}{\sqrt{2}}
\frac{\alpha }{2\pi \sin ^{2}\Theta _{W}}\lambda _{t}(\bar{d}s)
(\overline{\nu }_{j}\nu _{k})\widetilde{V}_{j\ell }^{\ast }
\widetilde{V}_{k\ell }X^{Dir}(x_{t},x_{\ell })+\mathrm{h}.\mathrm{c}.\,,
\end{equation}
\begin{equation}
\label{Xdirmix}
X^{Dir}(x_{t},x_{\ell })=\frac{m_{\nu }m_{s}}{M_{W}^{2}}\frac{x_{t}}{2}
\left[ \frac{1}{(x_{t}-1)^{2}}\left( 2-x_{t}+\ln (x_{t})\frac{x_{t}-3}
{2(x_{t}-1)}\right) +\frac{3}{16}\frac{M_{W}^{2}}{M_{H}^{2}}+O(x_{\ell })
\right]. 
\end{equation}
The unitarity of $\widetilde{V}$ and the smallness of the charged
lepton masses,  $x_{\ell }<10^{-3}$ , induce a very effective
leptonic GIM mechanism. Consequently, to a very good approximation, 
$X^{Dir}(x_{t},x_{\ell })$ is independent of $x_{\ell }$ and the sum 
over $\ell$ in (\ref{Hdirmix}) gives: 
\begin{equation}
{\mathcal{H}}_{mix}^{Dir}\cong \sum _{\ell }\frac{G_{F}}{\sqrt{2}}
\frac{\alpha }{2\pi \sin ^{2}\Theta _{W}}\lambda _{t}\, (\bar{d}s)
(\overline{\nu }_{\ell }\nu _{\ell })X^{Dir}(x_{t})
+\mathrm{h}.\mathrm{c}.\,.
\end{equation}
This is nothing but ${\mathcal{H}}_{CPC}^{Dir}$ of eq. (\ref{HDCPC}) 
which was derived without mixing. Thus the effects of lepton flavor 
mixing here are negligible.

\subsubsection{Majorana neutrinos}
The new contribution to the decay is generated by the $ \nu\nu HH$
vertex. Since this vertex is flavor diagonal (in the mass basis), 
lepton flavor mixing has no effect.

\subsection{V-A Operators}
\subsubsection{Introduction}

Without mixing, the $V-A$ operator creates the neutrino pair in a 
CP even state, thus requiring purely CPV contributions. 
With mixing the analysis is modified.
The effective Hamiltonian is modified as follows: 
\begin{equation}
\label{Hleadmix}
{\mathcal{H}}_{Lead}=\sum _{k,j,\ell }\frac{G_{F}}{\sqrt{2}}
\frac{\alpha }{2\pi \sin ^{2}\Theta _{W}}\lambda _{t}
\widetilde{V}_{j\ell }^{\ast }\widetilde{V}_{k\ell }
X(x_{t},x_{\ell })(\bar{s}d)_{V-A}(\overline{\nu }_{j}\nu _{k})_{V-A}
+\mathrm{h}.\mathrm{c}.\, ,
\end{equation}
where \cite{IL}: 
\begin{equation}
\label{X}
X(x,x_{\ell })\cong X_{0}(x,x_{\ell })=\frac{x}{8}\left[ \frac{x+2}{x-1}
+\frac{3x-6}{(x-1)^{2}}\ln x+O(x_{\ell })\right] \, .
\end{equation}
Neglecting the $O(x_{\ell })$ terms in (\ref{X}) leads to
the leading order CPV contribution \cite{IL}. However, if the 
$O(x_{\ell })$ contribution is not neglected, we find that 
the outgoing neutrinos do not necessarily carry the same flavor. 
Since (as seen above) the off diagonal contribution must be 
proportional to $x_{\ell }$, the electron contribution is negligible.
Moreover, in order to get an upper limit on the flavor mixed 
contributions, the dominance of the internal 
$\tau$ amplitude is assumed which means that:  
\begin{equation}
\label{assum}
\left | \frac{\widetilde{V}_{j\tau }^{\ast }\widetilde{V}_{k\tau}}
{\widetilde{V}_{j\mu }^{\ast }\widetilde{V}_{k\mu }}\right | 
> \frac{x_\mu}{x_\tau}
\sim \frac {1}{300} ~.
\end{equation}  
Consequently, ${\mathcal{H}}_{Lead}$ has the following off-diagonal 
part: 
\begin{equation}
\label{Hleadtau}
{\mathcal{H}}_{Lead}^{mix}\cong \sum _{k\neq j}\frac{G_{F}}{\sqrt{2}}
\frac{\alpha }{2\pi \sin ^{2}\Theta _{W}}\lambda _{t}
\widetilde{\lambda}_{jk}
\widetilde{X}_{0}(x_{t},x_{\tau })(\bar{s}d)_{V-A}
(\overline{\nu }_{j}\nu _{k})_{V-A}+\mathrm{h}.\mathrm{c}.\, ,
\end{equation}
where \( \widetilde{X}_{0} \) is the tau-mass dependent part of 
$X_{0}(x,x_{\ell })$ in (\ref{X}), and
$ \widetilde{\lambda}_{jk}\equiv \widetilde{V}_{j\tau }^{\ast }
\widetilde{V}_{k\tau }\, .$

\subsubsection{Dirac Neutrinos}
In the case of Dirac neutrinos the state 
$|\overline{\nu }_{j}\nu _{k}\rangle$ is not a CP eigenstate.
Then when ${\mathcal{H}}_{Lead}^{mix}$ of eq. (\ref{Hleadtau}) 
acts on the vacuum it creates the neutrinos in a mixed CP 
eigenstates.
Therefore it contaminates the SM leading decay products, which are 
generated in a CP even final state.
In order to find the magnitude of these new contributions, we  
calculate the decay rate, summed over all the 
final mixed flavor states:

\begin{equation}
\label{rate}
\Gamma ^{mix}\equiv \sum _{j\neq k} \Gamma _{
(K_{L}\rightarrow \pi ^{0}\overline{\nu} _{j}\nu _{k})}
=\sum _{j\neq k}\int 
\left| M(K_{L}\rightarrow \pi ^{0}\overline{\nu} _{j}\nu _{k})
\right| ^{2}d\Gamma \, ,
\end{equation}
where the matrix element $M$ in (\ref{rate}) is given by 
(using standard $V_{CKM}$ parameterization):
\begin{equation}
\label{finrate}
M(K_{L}\rightarrow \pi ^{0}\overline{\nu} _{j}\nu _{k})
\propto \Im\lambda _{t}
\widetilde{X}_{0}\langle \pi ^{0}|(\bar{s}d)_{V-A}|K^{0}\rangle \otimes
\langle \overline{\nu} _{j}\nu _{k}| \widetilde{\lambda}^{\ast }_{kj}
(\overline{\nu }_{k}\nu _{j})_{V-A}|0\rangle \, .
\end{equation}
Thus the upper bound on the rate is:
\begin{eqnarray}
\label{AleadCPC2}
\Gamma ^{mix}\propto 
\sum _{k\neq j} & \left| \Im\lambda _{t}  \widetilde{X}_{0}
\langle \pi ^{0}|(\bar{s}d)_{V-A}|K^{0}\rangle \, \langle 
\nu _{j}\overline{\nu }_{k}|(\overline{\nu }_{j}\nu _{k})_{V-A}|0\rangle 
\right| ^{2}\left| \widetilde{\lambda}_{jk}\right| ^{2} \, .
\end{eqnarray}
Note that: 
\begin{equation}
\label{Bound} 
\sum _{k\neq j} \left| \widetilde{\lambda}_{jk}\right| ^{2} 
\leq \frac{3}{4} \,. 
\end{equation}
We learn that the new contributions are suppressed by at least
$x_\tau^2={\cal O}(10^{-7})$. The decay is then still dominated 
by the leading SM CPV contribution  to an excellent approximation. 
Note, however, that the suppression here can be far milder than 
the one that we found for lepton flavor diagonal decays.
Furthermore, the interesting new contributions that lead to 
a CP odd final state arise from a double insertion of CP 
violating couplings. Therefore, all mixing related contributions 
vanish in the CP symmetry limit, similarly to the SM leading CPV 
contributions.

\subsubsection{Majorana Neutrinos}
For Majorana neutrinos we can rewrite the neutrinos $V-A$ operator 
as follows \cite{Kay}:
\begin{equation}
\label{Majv-a} 
\overline{\nu }_{j}\gamma ^{\mu} (1-\gamma ^5)\nu _{k}\propto
{\nu }^{T}_{j} \Omega \gamma ^{\mu} (1-\gamma ^5)\nu _{k}
\, ,
\end{equation}
where $\Omega$ is the charge conjugation matrix.
Moreover, for Majorana field $(\nu)^c= \nu $, which leads us to the 
following CP transformation law for the 
neutrinos $V-A$ operator: 
\begin{equation}
\label{CPlaw} 
CP\left\{ (\overline{\nu }_{j}
\nu _{k})_{V-A}\right\} CP^{-1} \propto (\overline{\nu }_{j}
\nu _{k})_{V-A}\, ,
\end{equation} 
for any $j,k$ \,.
This means that the neutrino pair is created in a CP even state.
Therefore the contribution here is singly CPV. 
The decay rate calculation is the same as in the case of 
the Dirac neutrinos (eq. \ref{rate}-\ref{AleadCPC2}) and therefore
yields the same negligible contribution (\ref{AleadCPC2})
(contributions suppressed by at least $x_{\tau}^2$).   
Nevertheless, the suppression here is by far smaller than the 
one that we found in (\ref{RGmCPCCPV}).

\subsection{Conclusions}
The effects of lepton flavor mixing can be summarized as follows:
\begin{itemize}
\item[(i)] The scalar operator CPC contribution is practically 
unchanged for both Dirac and Majorana neutrinos. 
\item[(ii)] The V-A operator leads to 
a new contributions all sensitive to the CPV sector of the theory. 
\item[(iii)]The ratio between the mixing contribution and the leading 
CPC contribution without mixing \cite{BI} is given by:
\begin{eqnarray}
\label{mixrate}
\sum _{k\neq j}\left( \frac{1}{a_{\chi}\lambda _c}
\frac{M_{W}^{2}} {m_{K}^{2}}
\right) ^{2}\frac{1}{{\mathrm{R}}_{\mathrm{kin}}}\left| 
\Im\lambda _{t}\widetilde{\lambda}_{kj}
\frac{m^{2}_{_{\tau }}}{M_{W}^{2}}\right| ^{2} &\approx& { \sum _{k\neq j}} 
\left(\frac{1}{{\mathrm{R}}_{\mathrm{kin}}a_{\chi }^{2}}  \right) \left| 
\widetilde{\lambda }_{kj}\frac{\Im \lambda _{t}}{\lambda _{c}}
\frac{m^{2}_{_{\tau }}}{m_{K}^{2}}\right| ^{2} 
\\
&\approx & \sum _{k\neq j}
0.1\left| \widetilde{\lambda }_{kj}\right| ^{2}<0.1\: ,
\nonumber
\end{eqnarray}
where $a_{\chi}={\mathcal{O}}\left( \frac{m_{K}^{2}}{8\pi ^{2}f^{2}}\right) 
\sim 0.2$ is a chiral suppression factor, and 
${\mathrm{R}}_{\mathrm{kin}} \sim 0.01$ is a phase space 
integration factor.
\end{itemize}
Since the mixing-related contribution are at most at the order of the 
SM CPC contribution, they are negligible and do not change
the present status of the $K_{L}\rightarrow \pi ^{0}\nu \bar{\nu}$ 
decay.

\section{ Final Conclusions}
In this work we examined the question of whether the SM CP violating 
contributions to the $K_{L}\rightarrow \pi ^{0}\nu \bar{\nu}$ decay
are still dominant in the presence of three additional
types of contributions:
\begin{itemize}
\item[(i)] CPC contributions related to massive Dirac neutrinos.
\item[(ii)] CPC contributions related to massive Majorana neutrinos.
\item[(iii)] Contributions related to flavor mixing effects.
\end{itemize}
We found an unambiguous answer to our question:
\begin{itemize}
\item[(i)] For the massive Dirac neutrinos contribution: 
\begin{equation}
\frac{\Gamma _{CPC}^{Dir }(K_{L}\rightarrow
\pi ^{0} \nu \bar{\nu})}{\Gamma _{CPV}(K_{L}
 \rightarrow \pi ^{0} \nu \bar{\nu})}\sim
\left( \frac{M_{K}}{M_{W}}\frac{m_{\nu }}{M_{W}}\right) ^{2}\lesssim
10^{-12}~ .  \label{RDir}
\end{equation}
\item[(ii)] For the massive Majorana neutrinos contribution: 
\begin{equation}
\frac{\Gamma _{CPC}^{Maj }(K_{L}\rightarrow
\pi ^{0} \nu \bar{\nu})}{\Gamma _{CPV} (K_{L}
 \rightarrow \pi ^{0} \nu \bar{\nu})}\sim
\left( \frac{\alpha }{64\sin ^{2}\Theta _{W}}\frac{m_{t}^{2}}{M_{H}^{2}}
\frac{m_{s}^{2}}{M_{W}^{2}}{\frac{M_{K}}{M_{W}}}\frac{m_{\nu }^{Maj}}{M_{W}}
\right) ^{2}\lesssim 10^{-29}~ .  \label{RMaj}
\end{equation}
\item[(iii)] For the flavor mixing effects contribution:
\begin{equation}
\frac{\Gamma ^{mix}(K_{L} \rightarrow
\pi ^{0} \nu \bar{\nu})}{\Gamma _{CPV}(K_{L}
\rightarrow \pi ^{0} \nu \bar{\nu})}\sim
\sum _{k \neq j}\left|\left(\frac{m_{\tau}}{M_{W}} \right)^2 
(\widetilde{\lambda}_{jk})\right| ^2 \lesssim 10^{-7} \, . 
\label{RMix}
\end{equation}
\end{itemize}
Moreover, if all neutrinos are lighter than $10~eV,$ then the 
$m_{\nu }$-related contributions are at least 24 (41) orders of 
magnitude smaller than the CP violating rate in the Dirac (Majorana) case.
It is clear then that the $K_{L}\rightarrow \pi ^{0}\nu \bar{\nu}$ decay
process provides a very clean measurement of fundamental, CP violating
properties and that it cannot probe neutrino masses.

The above results have interesting implications in the framework of
approximate CP. The SM picture of CP violation is not well tested. It could
be that $\varepsilon _{K}$ is small because CP is an approximate symmetry
and not because of the small SM mixing angles. This would require New
Physics to explain $\varepsilon _{K}$. For example, there exist
supersymmetric models with approximate CP \cite{YG,NW}. Generally, in such
models the CP violating phases fulfill $10^{-3}\lesssim \phi _{CP}\ll 1,$
where the lower bound comes from the experimental value of 
$\varepsilon _{K}$. Therefore, even if CP violation is accommodated by a 
source different from the CKM phases, we expect that 
$\Gamma (K_{L}\rightarrow \pi ^{0}\nu \bar{\nu})$ is no smaller than three 
orders of magnitude below the SM rate. Our study
implies that even in this extreme case, the CP violating contributions
still dominate.

\vspace{3mm}
{\large Acknowledgments} \vspace{3mm}

I thank Yossi Nir for his guidance in this work and 
Gerhard Buchalla for valuable comments on the manuscript.
I would also like to thank Gabriela Barenboim, Sven Bergmann and Galit 
Eyal for useful comments and discussions.

\end{document}